\documentclass[10pt,letterpaper]{article}
\usepackage[top=0.85in,left=2.75in,footskip=0.75in,marginparwidth=2in]{geometry}

\usepackage[utf8]{inputenc}
\usepackage{amsmath}
\usepackage{cite}
\usepackage{graphicx}
\usepackage{subcaption}
\usepackage{caption}
\usepackage{float}
\usepackage{geometry}
\usepackage[export]{adjustbox}
\usepackage{nameref,hyperref}

\usepackage{microtype}
\DisableLigatures[f]{encoding = *, family = * }

\raggedright
\usepackage{changepage}

\usepackage[aboveskip=1pt,labelfont=bf,labelsep=period,singlelinecheck=off]{caption}

\makeatletter
\renewcommand{\@biblabel}[1]{\quad#1.}
\makeatother

\usepackage{lastpage,fancyhdr,graphicx}
\usepackage{epstopdf}
\fancyheadoffset[L]{2.25in}
\fancyfootoffset[L]{2.25in}

\usepackage{color}

\definecolor{Gray}{gray}{.25}

\usepackage{graphicx}

\usepackage{sidecap}
\usepackage{amssymb}
\usepackage{wrapfig}
\usepackage[pscoord]{eso-pic}
\usepackage[fulladjust]{marginnote}
\reversemarginpar

\begin{document}
\vspace*{0.35in}

\begin{flushleft}
{\Large
\textbf\newline{Modelling Alzheimer's Protein Dynamics: A Data-Driven Integration of Stochastic  Methods, Machine Learning and Connectome Insights}
}
\newline
\\
Alec MacIver \textsuperscript{1,*},
Hina Shaheen  \textsuperscript{2}
\\
\bigskip
\bf{1} Department of Statistics, University of Manitoba, R3T 2N2, Winnipeg MB, Canada
\\
\bf{2} Faculty of Science, Department of Statistics, University of Manitoba, R3T 2N2, Winnipeg MB, Canada
\\
\bigskip
* macivera@myumanitoba.ca

\end{flushleft}

\section*{Abstract}
Alzheimer's disease (AD) is a complex neurodegenerative disorder characterized by the progressive accumulation of misfolded proteins, leading to cognitive decline. This study presents a novel stochastic modelling approach to simulate the propagation of these proteins within the brain. We employ a network diffusion model utilizing the Laplacian matrix derived from MRI data provided by the Human Connectome Project (https://braingraph.org/cms/). The deterministic model is extended by incorporating stochastic differential equations (SDEs) to account for inherent uncertainties in disease progression. Introducing stochastic components into the model allows for a more realistic simulation of the disease due to the multi-factorial nature of AD. By simulation, the model captures the variability in misfolded protein concentration across brain regions over time. Bayesian inference is a statistical method that uses prior beliefs and given data to model a posterior distribution for relevant parameter values. This allows us to better understand the impact of noise and external factors on AD progression. Deterministic results suggest that AD progresses at different speeds within each lobe of the brain, moreover, the frontal takes the longest to reach a perfect disease state. We find that in the presence of noise, the model never reaches a perfect disease state and the later years of AD are more unpredictable than earlier on in the disease. These results highlight the importance of integrating stochastic elements into deterministic models to achieve more realistic simulations, providing valuable insights for future studies on the dynamics of neurodegenerative diseases.

\section*{1. Introduction}

Alzheimer's disease (AD) is a neurological disorder that primarily affects cognitive function, typically manifesting in older adults and often characterized by dementia. It is estimated that approximately 416 million people suffer from AD globally \cite{globalstats}. Approximately 5.5 million Americans in 2017 live with AD \cite{factsAD}, with projections suggesting this number could rise to 13.8 million by 2050 \cite{factsAD}. In 2017, an individual in America develops Alzheimer's Disease every 66 seconds \cite{factsAD}. The disease's complex and multi-factorial nature, which involves a combination of genetic, environmental, and lifestyle factors, makes pinpointing its exact cause challenging. Furthermore, AD manifests differently across individuals, and it can develop silently for decades before clinical symptoms become apparent, complicating early diagnosis and intervention \cite{Silentdecades}. Ethical concerns also arise when conducting research on vulnerable populations, such as those with cognitive impairments, requiring stringent ethical considerations \cite{Ethics}.

A significant body of research has focused on misfolded proteins, particularly amyloid beta (A$\beta$) and tau ($\tau$), concerning AD \cite{MF1} \cite{MF2} \cite{MF3}. A$\beta$ is the result of the amyloid precursor protein (APP) through sequential enzymatic actions of $\beta$-secretase and $\gamma$-secretase \cite{misfoldedreview}. A$\beta$ monomers can aggregate to form oligomers, fibrils, and ultimately plaques, which are characteristic of AD pathology \cite{rolemonomers}. These A$\beta$ oligomers are highly toxic and disrupt synaptic function, leading to cognitive deficits \cite{oligomers}. In addition, $\tau$ a microtubule-associated protein, stabilizes microtubules in neurons \cite{taudescription}. In AD, tau becomes hyper-phosphorylated, reducing its ability to bind to microtubules and leading to the formation of neurofibrillary tangles (NFTs) \cite{misfoldedreview}. Evidence shows that A$\beta$ pathology precedes and exacerbates $\tau$ pathology by inducing $\tau$ hyper-phosphorylation and aggregation \cite{ABprecedes}. The interplay between A$\beta$ and $\tau$ is complex and critical for understanding the entire progression of AD.

Modelling AD propagation has evolved significantly, encompassing in vitro, in vivo, and in silico approaches. Early models primarily involved transgenic mice expressing human amyloid precursor protein (APP) and presenilin mutations to mimic amyloid beta (A$\beta$) plaque formation \cite{transgenicmice}. Subsequent advancements introduced tauopathies through $\tau$ transgenic mice, recapitulating neurofibrillary tangles \cite{tautransgenismice}. Recent models include human induced pluripotent stem cell (iPSC)-derived neurons and three-dimensional brain organoids, providing more human-relevant systems. These models have been instrumental in elucidating AD pathophysiology and testing therapeutic interventions despite some limitations in fully replicating the complexity of human AD.

The multi-factorial nature of AD \cite{carmo2013multifactorial} makes it difficult to model the disease. Many external factors like Diet \cite{Diet}, Exercise \cite{maci2012physical}, Sleep \cite{sleep} and Age \cite{age} are found to be related to AD. A stochastic approach accounts for the uncertainty regarding the multi-factorial nature of AD. Without some noise in the model, the results will always be the same which is an unrealistic assumption of how AD operates. This is why noise is important when modelling AD progression.

This study aims to model the propagation of misfolded proteins, adopting methods from recent advancements in the field \cite{fornari2019prion}. Additionally, this study seeks to incorporate a stochastic element to account for the inherent uncertainty in disease progression, thereby enhancing the model's robustness and predictive power. After the stochastic term has been added we run several simulations to study the impact of the new random process through exploratory data analysis and Bayesian inference to understand the distribution of relevant parameters.

\section*{2. Methods and Materials}

The paper is organized as follows, we define the adjacency matrix given the data from the human connectome project [Section 2.1]. An ordinary differential equation (ODE) is used to model how misfolded protein concentration increases in the brain over time [Section 2.2]. A stochastic differential equation (SDE) is used to add noise by the Weiner Process [Section 2.3]. After simulation, concentrations of misfolded protein are obtained at different times which are inferred at relevant time points using the approximate Bayesian computation method (ABC) [Section 2.4]. This section highlights AD modelling approaches including deterministic and stochastic network models. 

\subsection*{2.1 Data $\&$ Network Diffusion Model}

We use MRI data from the Human Connectome Project (HCP) \cite{hcp}, which has been reconstructed into an undirected graph $G$ with $N$ nodes/vertices and $E$ edges. The data contains 1064 brains each with 129 nodes and each with a varying number of edges. Denoted as follows,
\[
G = (V, E)
\]
\[
V(G) = \{v_1, v_2, \ldots, v_N\}
\]
\[
E(G) = \{\{v_i, v_j\} \mid v_i, v_j \in V \text{ and } v_i \neq v_j \}
\]
Each edge \( \{v_i, v_j\} \in E(G) \) connects the vertices \( v_i \) and \( v_j \).
\[
|V(G)| = N \quad and \quad |E(G)| = E
\]

The network diffusion model uses the undirected graph $G$ with $N$ vertices and $E$ edge defined above. This network diffusion model was chosen as it is a computationally effective way to visualize the human brain from the given data. Each edge is weighted which is defined as such in an adjacency matrix,
\[
A_{ij} = \frac{n_{ij}}{(l_{ij})^{2.5}} \qquad (1)
\]
where $n_{ij}$ are the number of fibers from $i$ to $j$ and $l_{ij}$ is the fiber length from node i to j \cite{shaheen2023astrocytic}. For $(l_{ij})^k$, $k$  is set equal to $2.5$ for computational efficiency. The network and adjacency matrix are averaged out over the 1064 brains via HCP and can be shown as such in Figure 1. This is an altered method adopted from \cite{shaheen2023astrocytic}. Figure 2 is a direct visualization of the adjacency matrix as a 3D graph, for better visualization the edges were omitted.

\begin{figure}[H]
\centering
\includegraphics[width=0.7\linewidth]{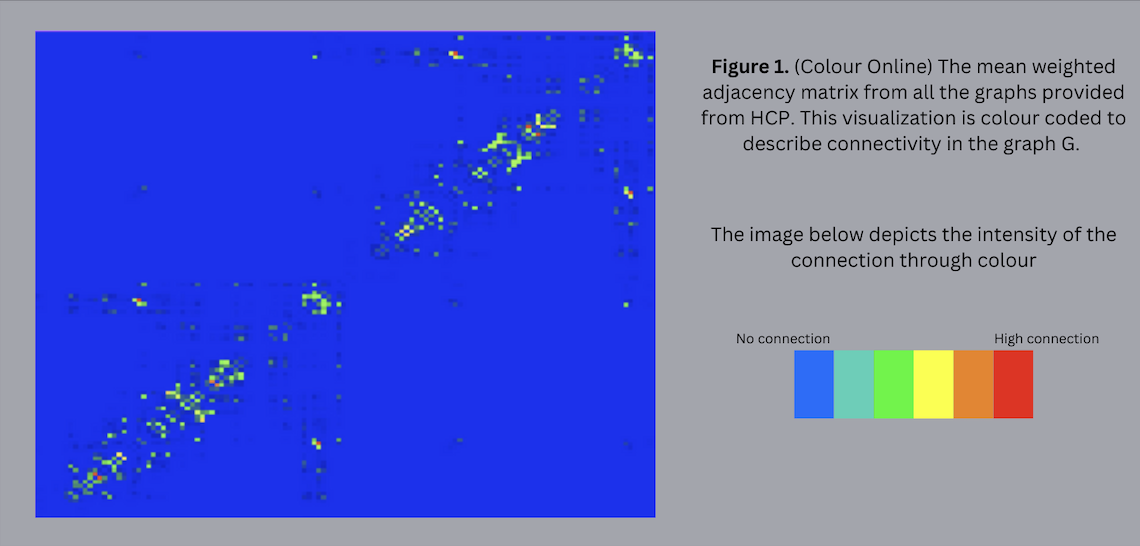}
\caption*{}
\label{fig:1}
\end{figure}

\begin{figure}[H]
\centering
\includegraphics[width=0.7\linewidth]{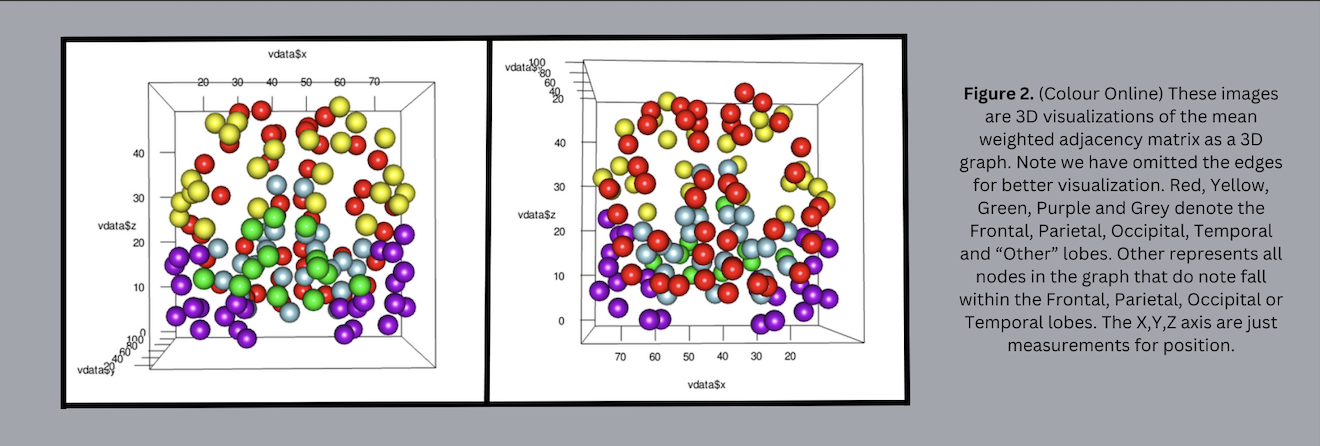}
\caption*{}
\label{fig:2}
\end{figure}

\subsection*{2.2 Fisher Kolomorgrov}
We expand the network diffusion model by incorporating the Fisher-Kolmogorov equation \cite{kolmogorov1937study} which characterizes misfolded protein aggregation as a nonlinear reaction-diffusion problem. This equation was chosen as it is commonly used to model a biological invasion process which fits well with our study \cite{whyfish}. The equation is denoted as follows,
\[
\frac{dc}{dt} = \nabla \cdot (D \cdot \nabla c + \alpha c [1 - c]) \qquad (2)
\]
Where $c\in[0,1]$ is the misfolded protein concentration and $\alpha$ is the conversion rate. The diffusion tensor D characterizes the speed and direction of misfolded protein spreading. We discretize equation (2) with the Laplacian matrix derived from our adjacency matrix. $L_{ij}$ is the Laplacian matrix which is found through the following formula \cite{biyikoglu2007laplacian},
\[
L_{ij} = D_{ij} - A_{ij} \qquad (3)
\]
$D_{ij}$ represents the degree matrix for the graph $G$, which is the matrix that is a diagonal matrix that represents the degree of each vertex. This can be derived via the following formula.
\[
 D_{ii} = \text{diag}\sum_{j=1, j \neq i}^{N} A_{ij} \qquad (4)
\]
This results in the following discretized equation adopted from a previous paper \cite{fornari2019prion} that will model the propagation of the misfolded proteins for $N$ unknowns, $i,j=1,2,...,N$.
\[
\frac{d c_i}{d t} = -\sum_{j=1}^{N} L_{ij} c_j + \alpha c_i[1 - c_i] \qquad (5)
\]
$N$ is the number of vertices in $G$, $L_{i,j}$ is the Laplacian matrix from equation (3), $\alpha$ is the conversion rate constant which will be set to $\alpha = 0.5$. A previous paper establishes the initial concentration of $c = 0.1$ in the entorhinal cortex \cite{fornari2019prion}. This ODE is solved numerically using the package desolver in Rstudio to approximate the ODE \cite{desolve2016}.  

\subsection*{2.3 Stochastic Term}
Expanding on the deterministic network diffusion model, we provide a novel approach to modelling misfolded protein propagation with noise. It is important to account for uncertainty, in biological models as many external or internal factors can affect the results of our observations. For example, it has been hypothesized that diet, vascular health and exercise are found to correlate with AD progression \cite{Diet} \cite{causesofAD}. A stochastic model allows for a realistic simulation that will often be more reliable than a deterministic model as the noise accounts for some of the unrecorded randomness that occurs within an individual's life. The noise term (or diffusion term) is defined as follows,
\[
\sigma^2c_idW(t) \qquad (6)
\]
Putting this into our ODE results in the following SDE.
\[
{d c_i} = (-\sum_{j=1}^{N} L_{ij} c_j + \alpha c_i[1 - c_i])dt + \sigma^2c_idW(t) \qquad (7)
\]
The first term is the drift term, and the second is the diffusion term. The $W(t)$, represents the standard Wiener process defined on a complete probability space $(\Omega,\mathcal{F},\mathcal{P})$ and $\sigma^2 > 0$, denotes the intensities of white noise \cite{ZHANG2020105347}. We will be setting $\sigma$ equal to 0.9 and 2 to explore the effect white noise has on the model, along with how the intensity of white noise impacts the results. $\sigma = 0.9, 2$ are chosen as 0.9 will result in a relatively lower variation, while $\sigma = 2$ will result in a much more intense variation in our results. The Wiener Process is chosen as it models Gaussian noise, this reflects real-world phenomena within the biological setting \cite{bionormal}. To solve this SDE, we apply the Euler-Maruyama method, which allows us to approximate the solution that would otherwise be difficult to solve \cite{oduselu2019review}. The following is a general example of how the Euler-Maruyama method is applied.

\[
c_{n+1} = c_{n}+(-\sum_{j=1}^{N} L_{ij} c_{n} + \alpha c_{n}[1 - c_{n}])\Delta t + \sigma^2c_{n}\sqrt{\Delta t}\mathbf{Z} \qquad (8)
\]
Where the random vector $\mathbf{Z}\in\mathbb{R}^{N}$ follows a standard normal distribution.

\subsection*{2.4 Approximate Bayesian Computation}
ABC is an approach that provides low-cost numerical solutions that are data-driven. Bayesian inference is a robust tool that uses prior knowledge in the form of a prior distribution to estimate posterior probabilities. In this case, we will be looking at the mean of the misfolded protein concentration at a given time $t$. The data will be the results gathered from running the SDE via simulation. The assumed prior for $\mu$ will be a standard normal distribution with mean zero and standard deviation of ten, $N(0,10)$, this uninformative prior, is great for parameter estimation \cite{shaheen2024data}. The assumed prior for $\sigma$ will be a half-Cauchy with a location parameter of 0 and a scale parameter of 2.5, $Half-Cauchy(0, 2.5)$, this is also an uninformative prior which is often used for scale parameters such as $\sigma$ \cite{halfcauchy}. Given that the data $\mathbf{X} = (x_1,x_2, ..., x_N)$ follows a normal distribution, this results in the following likelihood function \cite{gelman1995bayesian},
\[
\mathbb{P}(\mathbf{X} \mid \theta, t) = \prod_{i=1}^{N} \frac{1}{\sqrt{2\pi\sigma^2}}\exp\left(-\frac{(x_i - \hat{x})^2}{2\sigma^2} \right) \qquad (9)
\]
Using Bayes Theorem \cite{Koch1990}, the posterior then can be modelled as follows,
\[
\mathbb{P}(\theta, t \mid \mathbf{X}) \propto \mathbb{P}(\mathbf{X} \mid \theta, t)\mathbb{P}(\theta) \qquad (10)
\]
This distribution will be used to infer the true parameter value of the mean concentration at time $t$ and will be acquired via the Monte Carlo Markov Chain (MCMC) algorithm \cite{beaumont2019approximate}. This algorithm is commonly used to numerically approximate the posterior when the posterior is computationally expensive to derive. The MCMC will be run over 4 chains with 5000 tuning samples, and 5000 post-tuning samples each. After convergence of the posterior distributions, we draw 20000 posterior predictive samples, which allows us to quantify the uncertainty of the inferred parameter. A posterior distribution for $\mu$ will be simulated for each lobe, providing an insight into different lobe behaviour in the presence of noise.

The network diffusion model is a useful tool for simulating AD progression. The data from HCP provides an excellent topological map of the brain that describes connectivity. The Fisher-Kolmogorov equation describes how the misfolded proteins move throughout the brain. The stochastic addition to the network diffusion model allows the simulation to account for uncertainty, which will provide a more realistic simulation given AD's multi-factorial nature. After running multiple simulations and collecting relevant data the ABC method will be employed to infer the posterior distribution for the mean concentration at a given time $t$. The overall goal is to study the effects of introducing noise within the network diffusion model, the results of which are covered in section 3.

\section*{3 Results and Discussion}
This section will summarize the findings from the deterministic model, stochastic model and Bayesian inference. The deterministic model provides insights into a more controlled environment, its results will be explored here to directly compare it to the results with the noise term. Our novel approach adds noise via the Wiener Process, this accounts for external factors and allows us to simulate misfolded protein propagation more realistically. The stochastic results will be displayed in box plots at selective time points as well as 95$\%$ confidence interval plots over time $t$. Bayesian inference will provide a good understanding of the parameter distributions based on the data acquired from the simulations, these results will be displayed through their posterior distributions as well as some summary statistics. The results produced in this section will have many implications related to AD and create questions that will require further research. All results measure the percent concentration of the misfolded protein within the system. 

\subsection*{3.1 Deterministic Results}
As mentioned above the deterministic model provides insights into a more controlled environment which does not account for uncertainties. Equation (5) can be expressed through Figure 3, we can see that the entorhinal cortex propagates the protein faster than any other node, which makes sense as it is the area where the disease begins. On the other hand, the frontal pole takes the longest to propagate. Concerning time it seems most of the nodes clear 50 percent around 20 years, while after 35 years, the brain is almost fully converted to bad protein.

\renewcommand{\thefigure}{3}
\begin{figure}[H]
\centering
\includegraphics[width = 0.7\textwidth]{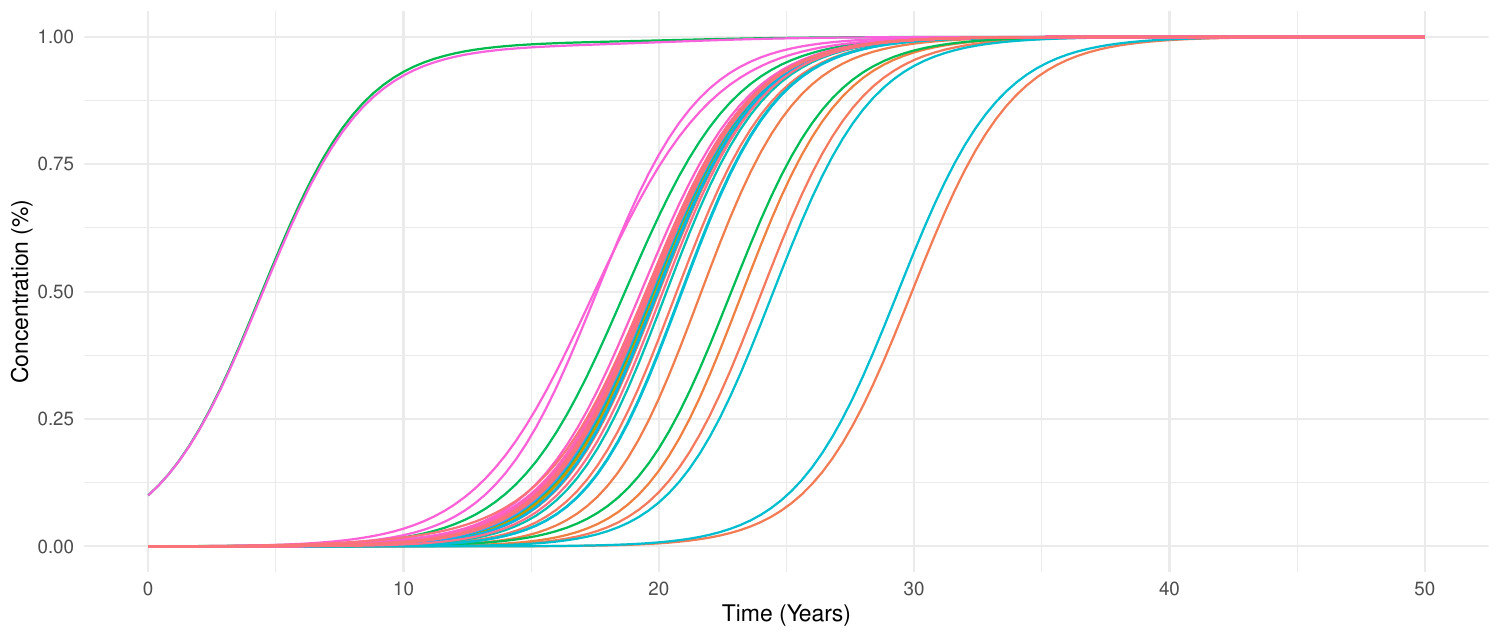}
\caption{(Colour Online) The solution to the ODE from equation (5). Each line represents a node in the brain network. This describes how MP propagates throughout our brains over time from a healthy state [c = 0] to a diseased state [c = 1]. Concentration (\%) is on the Y-axis, while Time (Years) is on the X-axis.}
\label{fig:3}
\end{figure}
The aggregate results for the deterministic model can be visualized in Figure 5. This was achieved by evaluating the mean concentration at each time point, $t$. The aggregate results follow similar trends to the individual nodes, such as after 20 years they clear 50 percent concentration. The disease state is reached approximately after 35 years and rapid growth is found after 15 years. In regards to different lobes, such as the Temporal, Frontal, Parietal and Occipital, Figure 4 visualizes how misfolded protein propagation behaves in different lobes. It is important to note that we denote all nodes that do not fall within the listed lobes as a group labelled "Other". Based on the results from Figure 4, it is clear that the Temporal lobe concentration percentage increases faster than the other lobes on the other hand the Frontal lobe appears to be the slowest, which is appropriate behaviour as found in previous studies \cite{frontallobe}. While Other, Occipital, Parietal and Other lobe regions increase at roughly the same speed.

Building on these deterministic results, our brain modelling reveals critical insights into the temporal and spatial progression of neurodegenerative diseases. The use of a coupled ODE system, as depicted in Figure 3, allows us to trace the propagation of misfolded proteins across different brain regions, demonstrating the entorhinal cortex's role as the primary site for rapid protein accumulation, consistent with clinical observations of disease onset. Moreover, by evaluating the mean concentration across all nodes (Figure 5), we can infer the overall brain's trajectory towards a diseased state, with clear phase transitions observed around 15 and 35 years. This model not only corroborates known patterns of disease spread but also provides a nuanced understanding of how different brain lobes, particularly the temporal and frontal lobes, contribute to the progression. Importantly, our results account for the inherent uncertainties in biological systems, ensuring that the predictions made by the model reflect realistic variability in disease propagation, thereby enhancing the model's applicability in understanding and potentially forecasting the dynamics of neurodegenerative diseases. Our results are validated with previous experimental and theoretical studies \cite{fornari2019prion} \cite{experimental_deterministic}. 

\renewcommand{\thefigure}{4}
\begin{figure}[H]
\centering
\includegraphics[width = 0.7\textwidth]{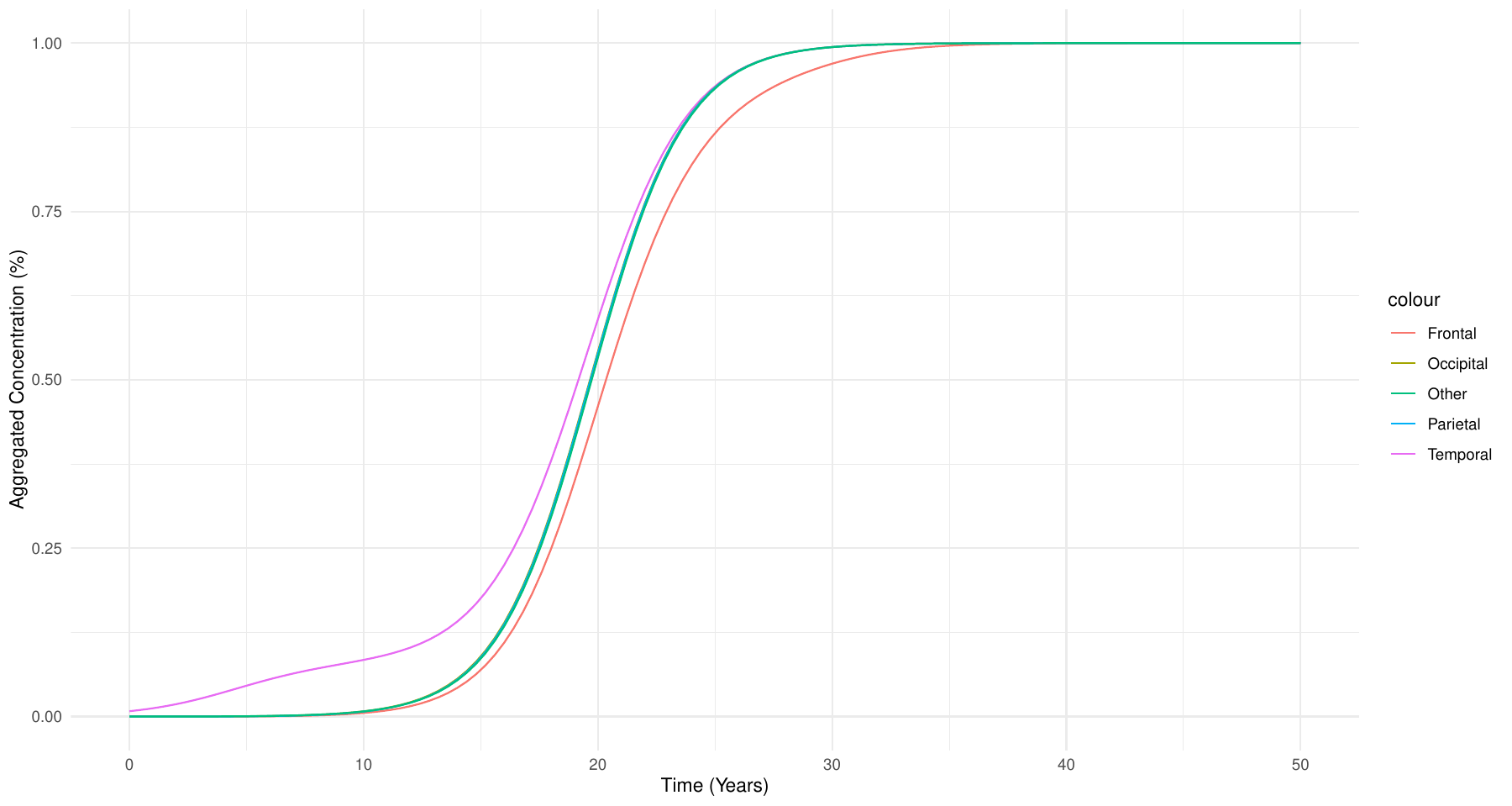}
\caption{(Colour Online) Focusing on individual lobes within the brain network. The Frontal, Occipital, Parietal and Temporal are visualized here. We also denote "Other" as all other nodes in $G = (V,E)$ that do not fall in any of the listed lobes. Concentration (\%) is on the Y-axis, while Time (Years) is on the X-axis.}
\label{fig:4}
\end{figure}

\renewcommand{\thefigure}{5}
\begin{figure}[H]
\centering
\includegraphics[width = 0.7\textwidth]{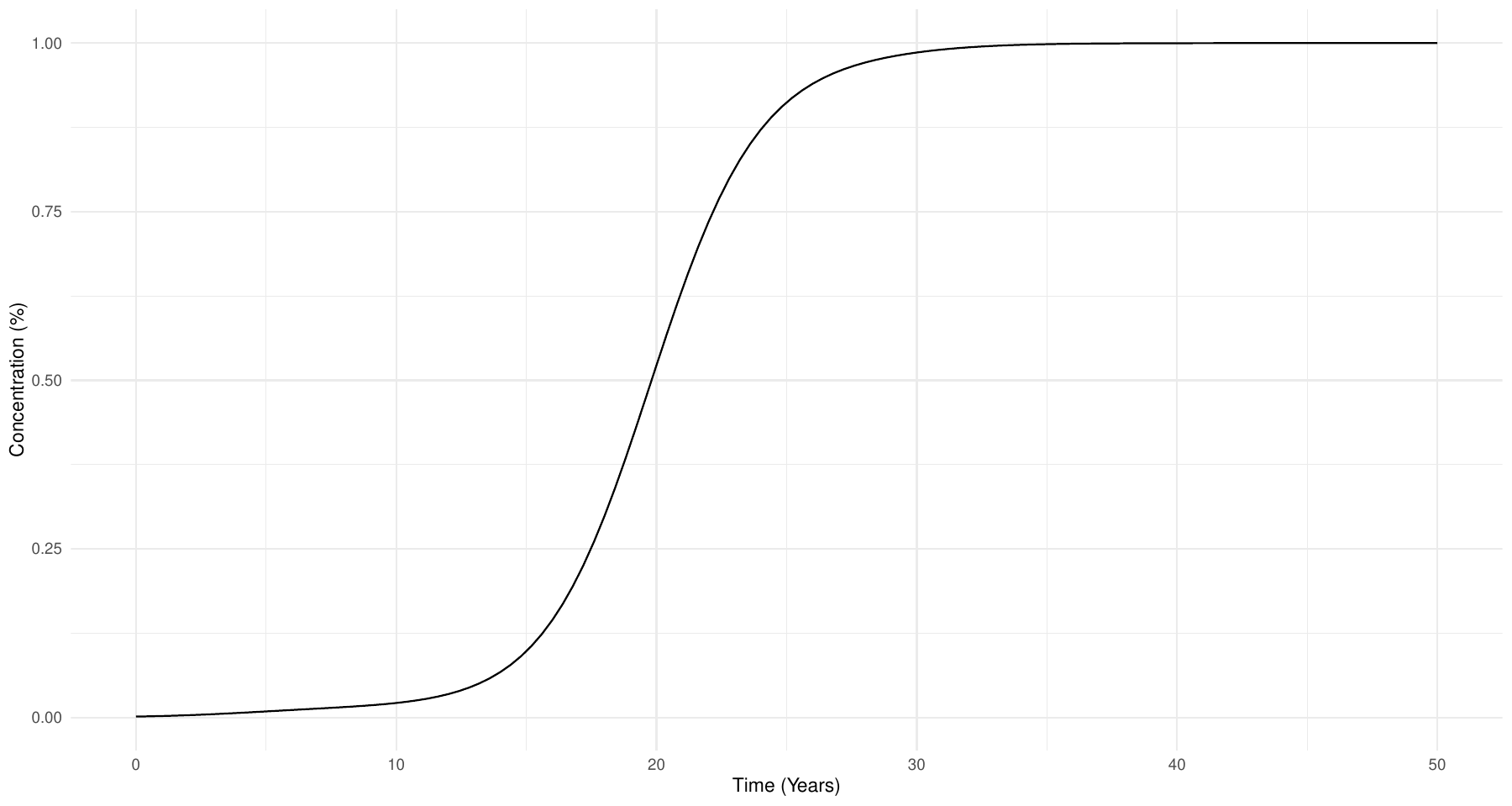}
\caption{(Colour Online) This is the aggregate ODE that looks at the mean concentration ($\%$) over time across all nodes. Concentration (\%) is on the Y-axis, while Time (Years) is on the X-axis.}
\label{fig:5}
\end{figure}

\subsection*{3.2 Stochastic Results}
The stochastic model provides insights into a more open environment, which accounts for uncertainties. We approximated Equation (7) via the Euler-Maruyama method, this section will cover the results found from running 1000 simulations to evaluate the effect of adding noise to the model. We ran 1000 simulations with $\sigma = 0.9$, and another 1000 with $\sigma = 2$. These values were chosen as such to compare what the simulation would look like with lower noise compared to higher noise. Figure 6 contains plots that display the difference in simulation variation when $\sigma = 0.9$ and $\sigma = 2$.

\begin{figure}[H]
\centering
\includegraphics[width = 0.9\textwidth]{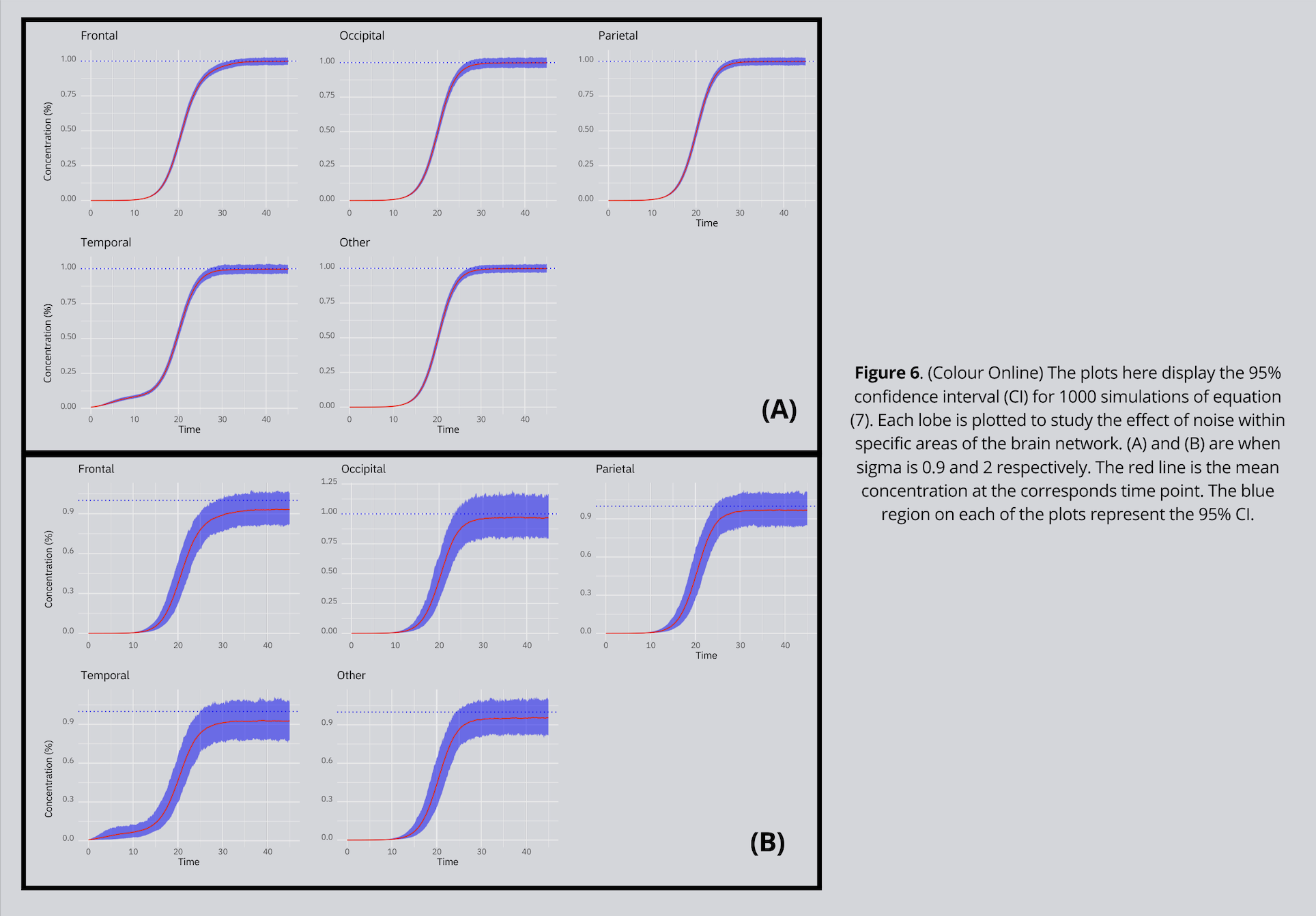}
\caption*{}
\label{fig:6}
\end{figure}

It is clear that when $\sigma = 0.9$ there is significantly less variation than when $\sigma = 2$, moreover, the mean concentration does not reach 1 when $\sigma = 2$. The variation seems to also be more intense after roughly 25 years. Regardless of noise intensity, it appears that there is little to no variation within the first 10 years of propagation. Figure 7 displays box plots of the concentration $c$ at varying times $t$ and $\sigma$. Based on all box plots the frontal lobe appears to fall behind the most typically in terms of concentration percentage. Moreover, their probability distributions were found to be approximately normal which is expected due to the Wiener process \cite{Wienertonormal}. The time point with the most influential observations is when $t = 30$, here the mean concentration is not always 1. The frontal lobe at $t = 30$ follows this pattern more aggressively than the other lobes. It is also worth mentioning at $t = 10$ the variation for all lobes except Temporal are minimal. 

Tables [1-6] contain the sample mean and sample standard deviation recorded for each lobe after running 1000 simulations. These parameters were chosen as the plots were found to be approximately normal and the mean and standard deviation are typically the parameters associated with a normal distribution \cite{weiss2006course}. The tables provide a non-visual way to understand how the intensity of noise impacts our results. Unsurprisingly, when $\sigma = 2$ the standard deviation is higher than when $\sigma = 0.9$. Similarly to Figure 7, the Temporal lobe typically has a higher mean while the frontal typically has a lower mean.

Incorporating stochastic modelling into the analysis of protein dynamics further enhances our understanding of how random fluctuations impact the spread of misfolded proteins across the brain. As observed, higher noise intensity $(\sigma = 2)$ leads to greater variation in protein concentration, particularly beyond the 25-year mark, while lower noise intensity $(\sigma = 0.9)$ results in more consistent outcomes across simulations. This stochastic framework reveals that the frontal lobe, which already lags in protein accumulation under deterministic conditions, exhibits even more pronounced delays when subjected to higher noise levels. The approximately normal distributions observed in the probability densities, as expected due to the underlying Wiener process, emphasize the significance of accounting for stochastic influences. The box plots (Figure 7) and statistical summaries (Tables 1-6) provide critical insights into how variability across different brain lobes changes over time, highlighting that while early propagation shows minimal variation, the later stages are heavily influenced by stochastic effects. Notably, at t = 30, the stochastic model underscores the vulnerability of the frontal lobe to noise, as its mean concentration often fails to reach unity, suggesting a complex interplay between deterministic progression and random perturbations. These are the novel results found in our study. This analysis is pivotal for understanding the real-world implications of neurodegenerative disease progression, where environmental and biological variability must be accounted for in predictive models, reinforcing our findings' robustness and relevance in a broader biomedical context. Our results on the stochastic modelling approach in AD progression are validated with previous experimental and theoretical studies \cite{Stocahstic_Theo} \cite{Stochastic_Exp}.

\begin{figure}[H]
\centering
\includegraphics[width = 0.8\textwidth]{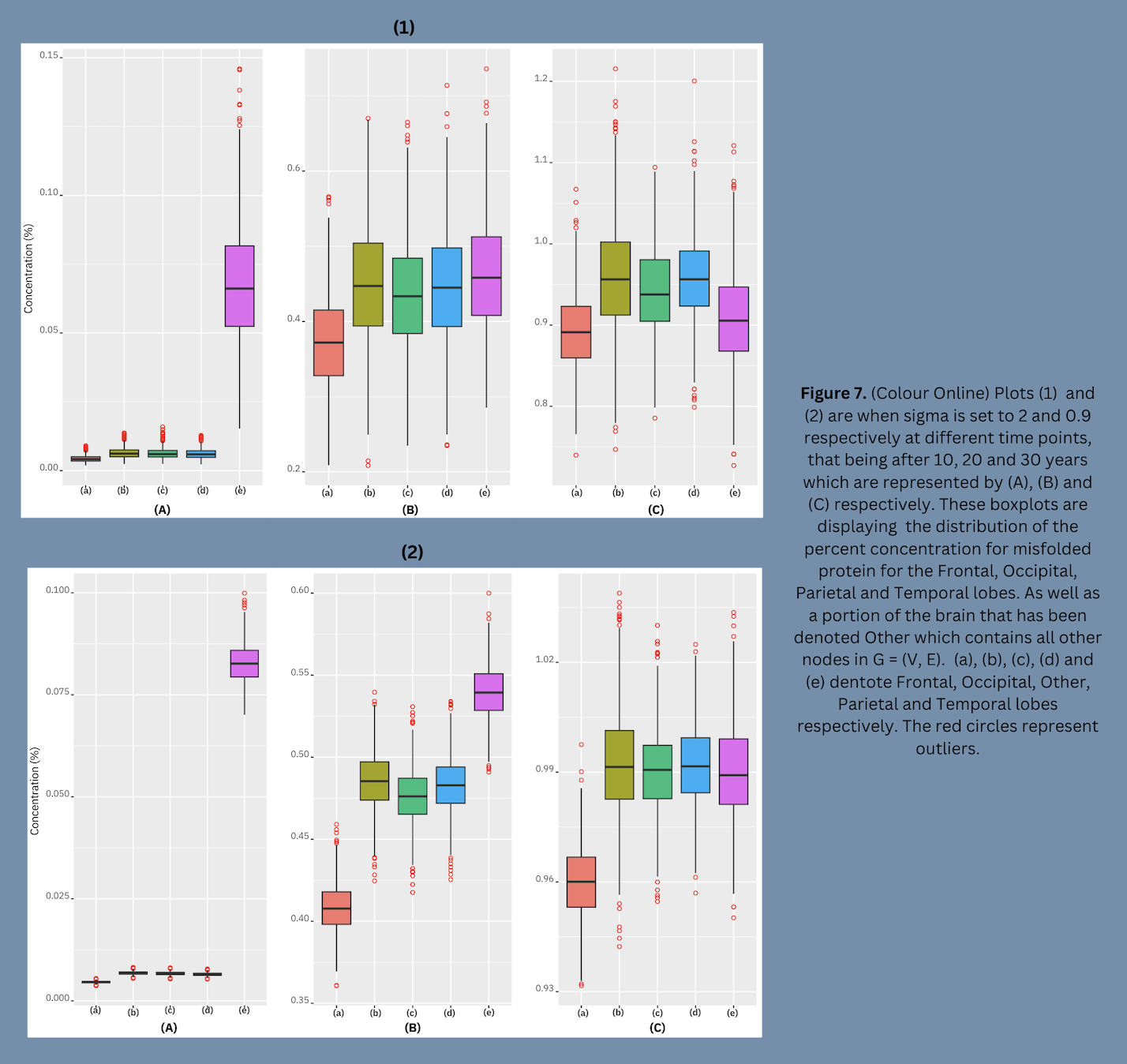}
\caption*{}
\label{fig:7}
\end{figure}

\begin{figure}[H]
\centering
\includegraphics[width = 0.8\textwidth]{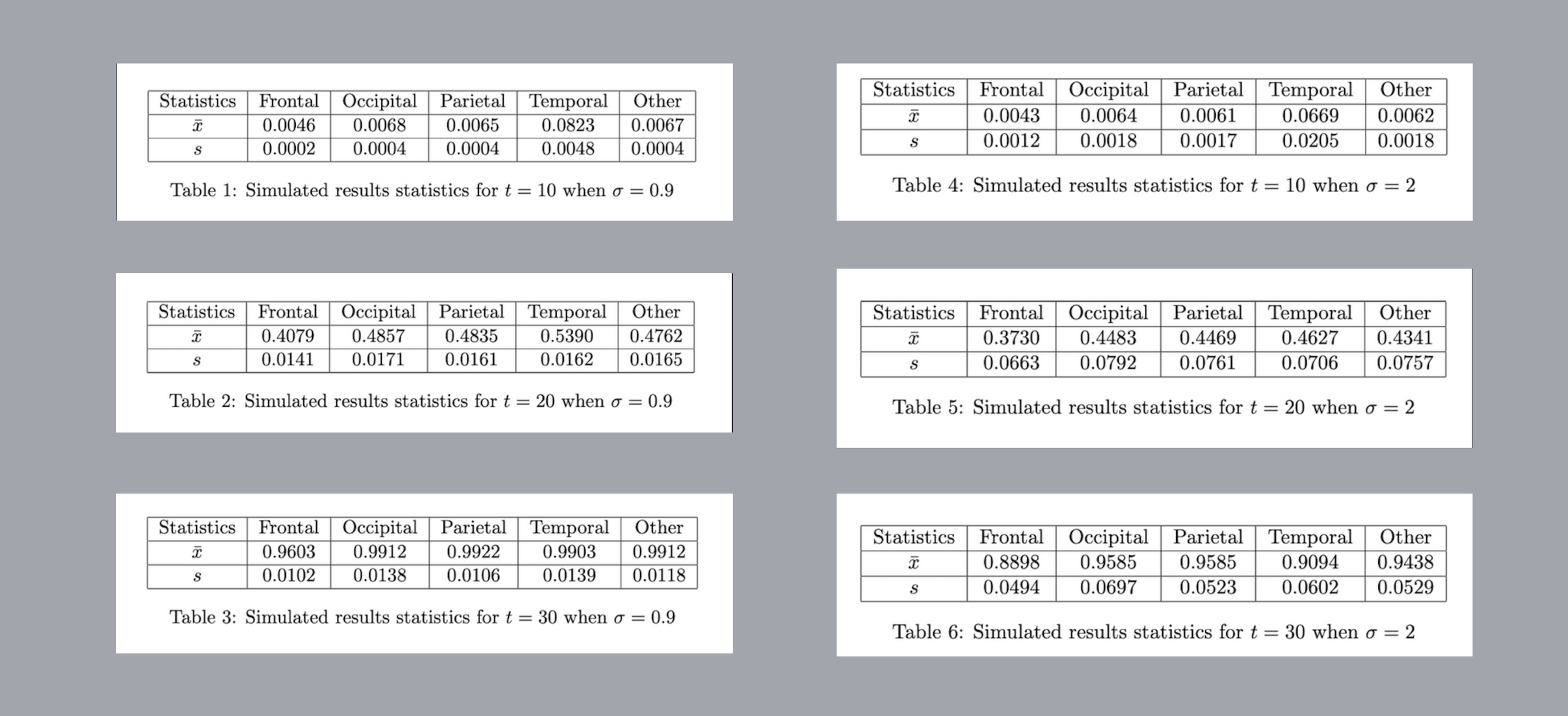}
\caption*{Tables: The mean and standard deviation for Figure 7 at a given time $t$ and noise intensity $\sigma$.}
\label{fig:a}
\end{figure}

\subsection*{3.3 Bayesian Inference}
Bayesian inference is important in modelling because it offers a flexible and principled approach to updating our understanding of a model based on new evidence. As mentioned in Section 2.4, the Prior used for the mean is a normal distribution with a mean of 0 and a standard deviation of 10. While the prior used for the standard deviation is a half-Cauchy distribution with a location parameter of 0 and a scale parameter of 2.5. Combined with the likelihood function, we derive the Posterior distribution for the parameter $\theta$. However, we use the ABC method to numerically approximate the Posterior via the MCMC algorithm. In which case, $\theta$ is the mean for percent concentration at a time $t$. Figure 7 represents the Posterior distributions at each time point $t = 10, 20, 30$. 

Bayesian Inference is conducted on the simulations where $\sigma = 2$ as the results are more interesting with more variation and the results when $\sigma = 0.9$ do not tell much of a story. Unsurprisingly, the posterior distributions are approximately normal given our prior and likelihood functions. Moreover, the posterior distributions for $\theta = \mu$ have low variation and are typically clustered around the mean. This suggests that it is common to see the same mean value after running multiple simulations. Tables 7, 8 and 9 represent some summary statistics at different time points that describe the posterior distributions in Figure 8. Comparing our findings from the posterior distributions to the results from Section 3.2 we can see that the mean concentration of the simulated values where $\sigma = 2$ is similar to the mean of the posterior distributions. The main findings within this Section are that the posterior distributions for the true mean percent concentration value at different time points $t$ are approximately normal. The standard deviation of these distributions is low and the mean of these distributions is close to their counterparts in Tables 4 - 6. The minimum and maximum explain the range for $\theta = \mu$'s distribution given $t$ and based on the results the range is small and since these are normal distributions the mean is the most likely value for $\theta = \mu$. These posterior distributions allow us to better understand the true parameter value for the mean percent concentration when $\sigma = 2$ at varying times $t$.

\begin{figure}[H]
\centering
\includegraphics[width = 1\textwidth]{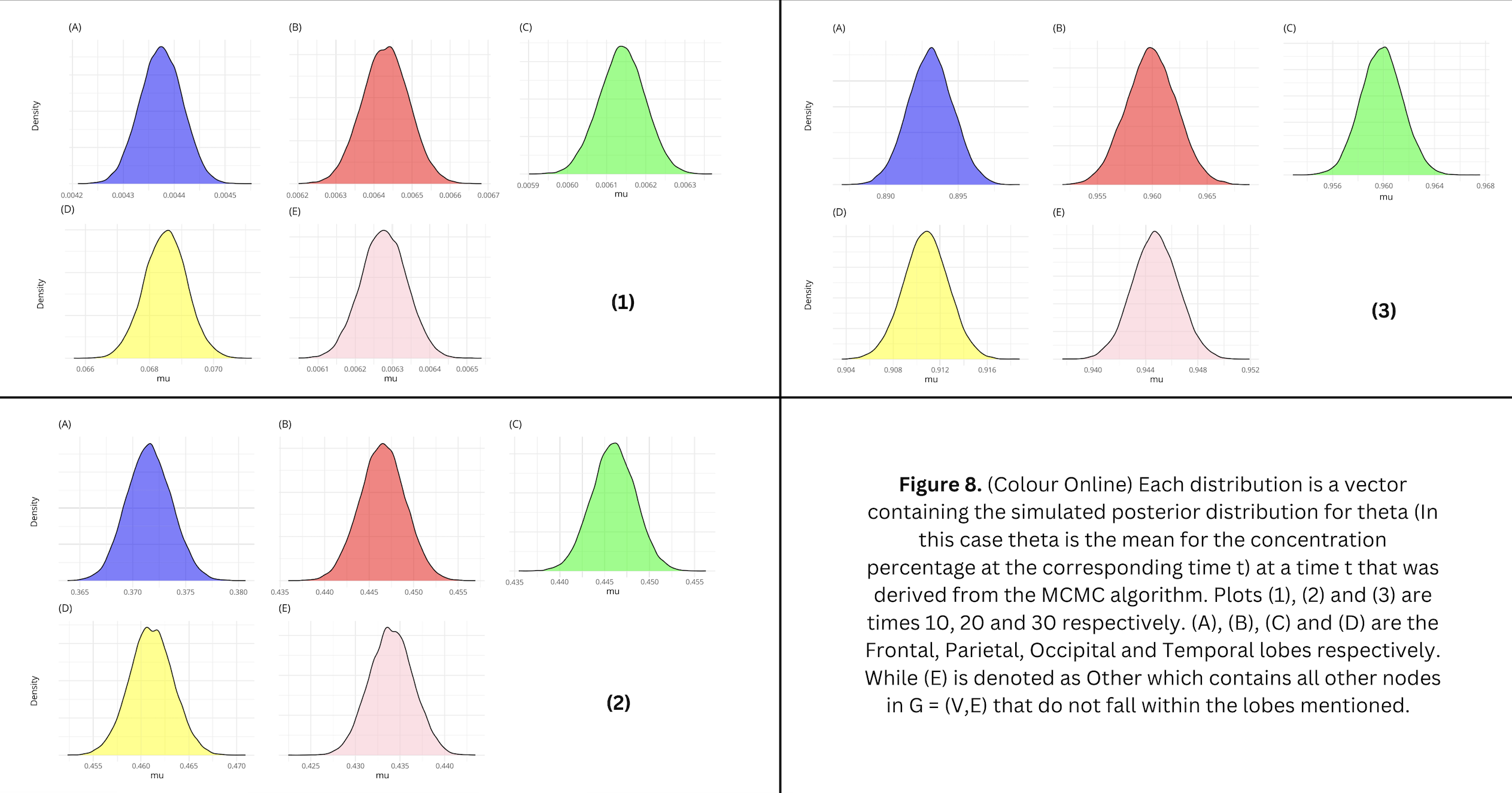}
\caption*{}
\label{fig:8}
\end{figure}

\begin{figure}[H]
\centering
\includegraphics[width = 0.7\textwidth]{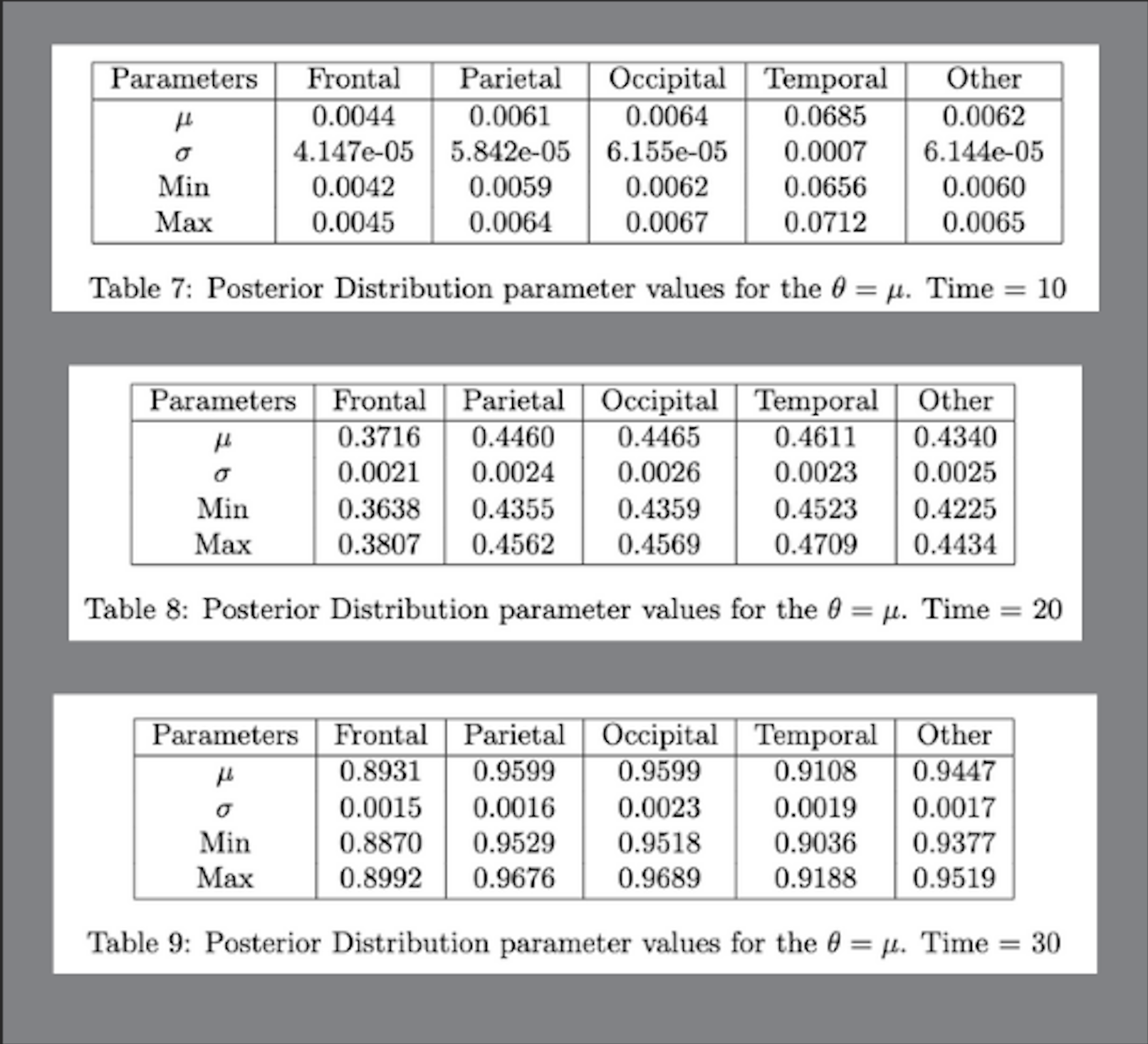}
\caption*{Tables: Parameters for Figure 8 at a given time $t$.}
\label{fig:b}
\end{figure}

\subsection*{3.4 Discussion}

The objective of this study is to model AD progression in the presence of noise. Our novel approach of incorporating the Weiner Process into the discretized Fish-Kolmogorov model has provided some insights into how AD misfolded protein propagation behaves. We find that typically the brain will not reach a perfect disease state (c = 1) in the presence of noise. Moreover, the most variation is present in the later years of AD which is 20 to 40 years after initial concentration. Little to no variation is present in the first 10 years after initial concentration. Based on the posterior distributions derived for the mean misfolded protein concertation after 10, 20 and 30 years we can see that they all approximately follow a normal distribution with relatively low variation. Deterministic results suggest that the frontal lobe lags behind the other lobes in terms of misfolded protein propagation speed.

Some of these results display the multi-factorial nature of AD. It suggests that AD can never truly reach a perfect disease state, moreover, the later years of AD appear to be unpredictable based on our model. It also suggests that things like motor function and higher-level cognitive functions are the last to be affected by AD as they reside in the frontal lobe. Some theories such as Network Degeneration Hypothesis \cite{drzezga2018network}, Vulnerable Neuron Hypothesis \cite{wang2020selective} and Prion-like Propagation Hypothesis \cite{prionlikepropegation} support the differences in lobe misfolded protein propagation speed over time. The stochastic results support theories related to the multi-factorial nature of AD and how environmental factors can influence misfolded protein propagation. Things like inflammation and immune response \cite{inflammation}, oxidative stress \cite{oxidativestress}, diet \cite{Diet}, sleep disturbances \cite{sleep} and physical activity/cognitive stimulation \cite{maci2012physical} are all hypothesized to impact misfolded protein propagation. The increased variation in the later years of AD could be related to the patient's age in the sense that as someone gets older, we have a more difficult time predicting misfolded protein behaviour. It is typically theorized that as someone increases in age they are more likely to experience AD which matches results that suggest variation increases as time progresses \cite{shaheen2024data}. As for applicable uses of the results found in the study, we are unfortunately limited through the computational study. One could use the results to open questions and investigations for further studies. A direct application would be to use the model to simulate a patient's AD progression based on their derived adjacency matrix from perhaps an MRI scan.

Looking at the deterministic results our findings are similar to previous studies as misfolded protein concentration reaches a disease state (c = 1) after roughly 40 years \cite{fornari2019prion}. This is interesting as previous studies use a different number of nodes. As for the novel stochastic approach, results found match well with previous studies regarding more uncertainty in the later years of AD after initial concentration \cite{AnotherLU}, \cite{3rdLU}. In terms of a perfect disease state (c = 1) in the presence of noise, the results for our model display that on average the perfect disease state (c = 1) will not be achieved on average. This suggests that environmental factors and inherent noise can prevent the disease from fully taking over the brain, which could have significant implications for therapeutic strategies and our understanding of disease resilience.

The number of limitations in this study where minimal but may have had large impacts. The overgeneralization of c as misfolded proteins instead of individually modelling $A\beta$ and $\tau$ may have been disadvantageous to understanding the true behaviour of the misfolded protein concentration. The use of uninformative priors may have also been too general for simulating the posterior distribution and could have impacted results negatively. In future studies, one could conduct inference on the mean weighted adjacency matrix to understand the probability distribution that the data from HCP follows, which surely would lead to more accurate results. Exploring models that incorporate $A\beta$ and $\tau$ and introducing noise to them would be interesting to study the results. Trying different priors and studying how the posterior distributions change would also surely give worthwhile results.

\section*{4 Conclusion}
In conclusion, this research highlights the importance of noise when modelling multi-factorial diseases. AD affects many people worldwide today and as mentioned previously in the study we can only expect that to increase as time moves on. The objective of this study is to develop a model for AD progression that accounts for uncertainty and to study the effects of introducing noise into the network diffusion model. This network diffusion model simulates connectivity and AD progression based on data from HCP. In the context of AD, a lot of information can be missed when simulating misfolded protein propagation with a deterministic model. Our novel approach of incorporating the Wiener Process into the network diffusion model introduces noise which creates an SDE, allowing for a more accurate simulation of AD progression. Results found reflect the real-world behaviour of AD. This can be seen with more uncertainty in the later years of the disease which results in the network typically not reaching a perfect disease state. We also find that the posterior distributions derived from Section 2.4 present insights on the true parameter distribution for $\mu$, which is a normal distribution with low variance. This suggests that we can expect to typically see common parameter values after many simulations. These results have significant implications for therapeutic strategies and our understanding of disease resilience. Future studies will be based on incorporating more sophisticated stochastic processes and personalized data into the models to enhance their predictive power and applicability. Specifically, by integrating additional biological variables such as varying protein clearance rates, differential inter-lobe connectivity, and individual patient data, these models can be tailored to reflect the unique progression of neurodegenerative diseases in different individuals. Furthermore, future research could explore the impact of various therapeutic interventions within this enhanced modelling framework, offering insights into the timing and effectiveness of potential treatments. The ultimate goal is to refine these models into robust tools that deepen our understanding of disease mechanisms and provide practical guidance in clinical decision-making, particularly in the early detection and personalized treatment of neurodegenerative disorders. Overall, refining the stochastic models to include additional biological factors, such as varying levels of protein clearance and inter-lobe connectivity, could provide a more nuanced understanding of disease progression. Moreover, incorporating patient-specific data could lead to personalized models that better predict individual disease trajectories. Additionally, exploring the effects of therapeutic interventions within this framework could offer valuable insights into the potential efficacy and timing of treatments to slow or halt the progression of neurodegenerative diseases. Further experimental validation of the model's predictions, particularly about the impact of noise on different brain regions, would also be beneficial in corroborating these findings and enhancing their applicability in clinical practice. Ultimately, this study lays the groundwork for a more integrative approach to modelling brain disorders, combining mathematical rigour with biological realism to advance our theoretical understanding and practical capabilities in managing these complex conditions.

\bibliography{library}

\bibliographystyle{abbrv}

\end{document}